# Effect of Pr doping on the superconductivity and interlayer coupling of $Bi_2Sr_{2-x}Pr_xCa_1Cu_2O_y$ system

H. Salamati [a],*   P. Kameli [a]   F. S. Razavi [b]

[a] Department of Physics, Isfahan University of Technology, Isfahan 84154, Iran
[b] Department of physics, Brock University, St. Catharines, Canada ON L2S 3A1

**Abstract**

The effect of Pr substitution on the superconductivity and interlayer coupling of $Bi_2Sr_2Ca_1Cu_2O_y$ system is investigated. Magnetic and transport measurements were performed for purposes of characterization. The superconducting transition temperature $T_c$ first increased and then decreased till it became zero at x=0.6. The effective superconducting volume also decreases due to Pr substitution. From the fluctuation conductivity analyses it is found that the interlayer coupling constant J decreases monotonically with increase of the Pr content. This result shows that the Pr doping weakens the $CuO_2$ interlayer coupling of Bi2212 system due to the loss of local superconductivity in the $CuO_2$ layers.

Keywords: High-$T_c$ superconductors, Substitution, Fluctuation conductivity, Interlayer coupling

## 1. Introduction

The study of the effect of impurities on the physical and superconducting properties of high-$T_c$ superconductors has long been recognized as being of great importance.

* Corresponding author. E-mail address: salamati@cc.iut.ac.ir

For $Bi_2Sr_2Ca_1Cu_2O_y$ (Bi2212) there have been extensive studies on the cationic substitution of Ca and Sr by the rare earth elements to change the carrier density and to probe the underlying mechanism of superconductivity[1-7].Some people believed that doping of Bi2212 with rare earth elements would lead to a depression of $T_c$ as a result of hole filling rather than Abrikosove-Gorkove (AG) pair breaking mechanism[5-7].In spite of extensive studies of Bi2212 system, in many ways, this material provide a superior medium. These studies have primary indicate the trends of $T_c$ versus doping concentration and did not include more extensive studies of magnetic, transport and thermodynamic behaviours. It is obvious that the interlayer coupling between superconducting $CuO_2$ layers play an important role in determining the electromagnetic and thermodynamic properties of high-$T_c$ superconductors. In Bi2212 system, due to the short coherence length and week interlayer coupling between $CuO_2$ layers, fluctuation conductivity plays a crucial role. Previous results showed that the fluctuation conductivity is influenced by doping element and also by irradiation [8,9]. From the irradiated samples it is seen that the interlayer coupling weakens because of distortions created in the $CuO_2$ layers. The Pr substitution in Bi2212 system is also expected to affect the fluctuation conductivity and interlayer coupling through a distortions in the $CuO_2$ bilayers. In this paper, we report the study on the effect of Pr doping on the superconductivity and interlayer coupling of $Bi_2Sr_{2-x}Pr_xCa_1Cu_2O_y$.

## 2. Experiment

Polycrystalline samples of $Bi_2Sr_{2-x}Pr_xCa_1Cu_2O_y$ were prepared by conventional solid state reaction method[7].The Dc magnetic susceptibility measurements were carried out using a Dc Superconducting Quantum Interference Device (SQUID) magnetometer in an applied magnetic field of 5G. The resistivity measurements were carried out on a bare shaped sample using closed-cycle refrigerator.

## 3. Results and discussion

The temperature dependences of zero field cooled Dc magnetic susceptibility of the samples are shown in Fig.1. As one clearly can see the onset of superconducting transition temperature ($T_c$) has a slight increase at first reaching a maximum at x=0.1 and then drops gradually. So x=0.1 must be near the optimal doping, while x=0 is in the over doped and x>0.1 samples are in the under doped regime. This suppression is not very rapid but, $T_c$ reaches zero suddenly when Pr content reaches about x=0.6.

According to the AG pair breaking phenomena[10] doping of superconductors with magnetic impurities causes a strong suppression of $T_c$. In this theory $T_c$ decreases linearly with the concentration of magnetic ions(x) for small values of x. While at a large value of x, the decrease in $T_c$ is more rapid. As shown in Fig.2 in our results $T_c$ suppression with x clearly deviates from the AG pair breaking law, which means that the magnetic nature of the Pr ions dose not play an important role in the mechanism of the $T_c$ suppression. The similarity in $T_c$ versus x for Pr in our measurements and La for the work of Ruan et al [4] suggest that both ions enter the lattice as trivalent ions and that there must be minimal exchange coupling between the rare-earth ions and the conduction holes within the $CuO_2$ layers. If the AG pair breaking phenomena was a significant factor in the suppression of $T_c$ in these materials, it would be expected that the concentration of dopants required for the observed suppression of $T_c$ would differ greatly in two cases, because of the substantial difference in their paramagnetic moments and their total angular momentum. This result suggest that the principal mechanism of suppression is rather a nonmagnetic charging effect due to the extra electron introduced at the Sr site by dopant i.e., the contribution to $T_c$ suppression mainly comes from the hole-filling effect.

As shown in Fig.1, the diamagnetic response signal decreases with increasing Pr concentration. The decrease of diamagnetic magnitude with the increase of Pr concentration may relate to the destruction of $CuO_2$ interlayer coupling induced by substituting Pr for Sr. In Pr doped system, The local superconductivity is lost in the vicinity of randomly distributed Pr ions. Therefore, the effective superconducting area decreases due to Pr substitution. The loss of local superconductivity in $CuO_2$ bilayers will dramatically weaken the interlayer coupling of the Bi2212 system.

In order to study the effect of Pr doping on the $CuO_2$ interlayer coupling of Bi2212, the fluctuation conductivity of the samples were studied in the framework of Aslamazov – Larkin (AL) Theory [11]. According to the AL theory, the fluctuation conductivity ($\Delta\sigma$) is defined as:

$$\Delta\sigma = A\varepsilon^{-\lambda} \qquad (1)$$

Where $\varepsilon = \dfrac{T-T_c}{T_c}$ is the reduced temperature in which $T_c$ is the mean field critical temperature and A is $\dfrac{e^2}{16\hbar d}$ and $\dfrac{e^2}{32\hbar \xi_c(0)}$ for 2D and 3D respectively. $\xi_c(0)$ is the coherence length along the c-axis and d is the effective interlayer distance between the neighbouring $CuO_2$ planes. The exponent $\lambda$ is –0.5 for the 3D fluctuation and –1 for the 2D fluctuations. The experimental value of $\Delta\sigma$ is found using the relation:

$$\Delta\sigma = \dfrac{1}{\rho(T)} - \dfrac{1}{\rho_n(T)} \qquad (2)$$

Where the $\rho(T)$ is the measured value of resistivity and $\rho_n(T)$ is the linearly extrapolated normal state resistivity. For strongly an isotropic superconductors Lawrence and Doniach [12] modified the AL theory by introducing a small coupling J for the neighbouring $CuO_2$ planes, so that the fluctuation conductivity becomes:

$$\Delta\sigma = \dfrac{e^2}{16\hbar d\varepsilon}(1+\dfrac{4J}{\varepsilon})^{\dfrac{-1}{2}} \qquad (3)$$

Where $J = \left[\dfrac{\xi_c(0)}{d}\right]^2$. This equation predicts a cross over from 2D ($\Delta\sigma \approx \dfrac{1}{\varepsilon}$) to 3D ($\Delta\sigma \approx \dfrac{1}{\varepsilon^{\frac{1}{2}}}$) behaviour and it occur at a temperature $T_0$ where $\varepsilon_0 = \dfrac{T_0}{T_c} - 1 = 4J$.

Fig.3. shows the temperature dependence of resistivity of pure and Pr doped samples. In Fig.4. the temperature derivative $\dfrac{d\rho}{dT}$, is plotted in the transition region of the samples. The peak of these curves is taken as $T_c$. See Table 1.

Fig.5.shows the logarithm of the fluctuation conductivity Ln ($\Delta\sigma$) vs. logarithm of reduced temperature Ln ($\varepsilon$) for pure and three doped samples. We notice that all of the plots manifest a crossover from the 3D fluctuation conductivity with critical exponents, $\lambda_1 \approx -\frac{1}{2}$, to the 2D fluctuation conductivity with critical exponents, $\lambda_2 \approx -1$.

The reduced crossover temperature $\varepsilon_0, \lambda_1, \lambda_2$ is listed in Table 1. These observations suggest that Pr doping do not influence the fluctuation conductivity analyses and that observed properties are intrinsic. The reduced crossover temperature $\varepsilon_0$ decreases with increasing the doping concentration x. The values of interlayer coupling constant J can be obtained from the reduced crossover temperature $\varepsilon_0$. Although the superconductivity of these samples changes from over doping (x=0) to the optimal doping (x=0.1) and then to under doping (x=0.2,0.4), the interlayer coupling constant J decreases monotonically with increase of Pr content as shown in Table 1. These results show that the Pr doping weakens the $CuO_2$ interlayer coupling of the Bi2212 system.

## 4. Conclusion

We report on the effect of Pr substitution on the superconductivity and interlayer coupling of $Bi_2Sr_{2-x}Pr_xCa_1Cu_2O_y$ system. The superconducting transition temperature $T_c$ has slight increase at first reaching a maximum at x=0.1 and then drops gradually. This suppression is not very rapid but, $T_c$ reaches zero suddenly when Pr content reaches about x=0.6. We believe that hole filling gives the explanation of $T_c$ suppression. The effective superconducting volume also decreases with Pr content. Fluctuation conductivity analyses show that the interlayer coupling constant J decreases with increase of the Pr content. This is may be due to the loss of local superconductivity in the $CuO_2$ layers.

## Acknowledgments

We would like to thank Isfahan University of Technology and Brock University for supporting this project.

**Figure captions**

Fig.1. Zero field cooled dc magnetic succeptibility of $Bi_2Sr_{2-x}Pr_xCa_1Cu_2O_y$ for different values of x at 5G.

Fig.2. Superconducting transition temperature $T_c$ versus Pr content x for the $Bi_2Sr_{2-x}Pr_xCa_1Cu_2O_y$ samples.

Fig.3. Resistivities of $Bi_2Sr_{2-x}Pr_xCa_1Cu_2O_y$ for different values of x.

Fig.4. Temperature derivative of the resistivity near $T_c$ for the $Bi_2Sr_{2-x}Pr_xCa_1Cu_2O_y$ samples.

Fig.5. Plots of the logarithmic fluctuation conductivity Ln ($\Delta\sigma$) versus logarithmic reduced temperature Ln ($\varepsilon$) for $Bi_2Sr_{2-x}Pr_xCa_1Cu_2O_y$ samples.

Table.1. Characteristic parameters of $Bi_2Sr_{2-x}Pr_xCa_1Cu_2O_y$ samples for different values of x.
.

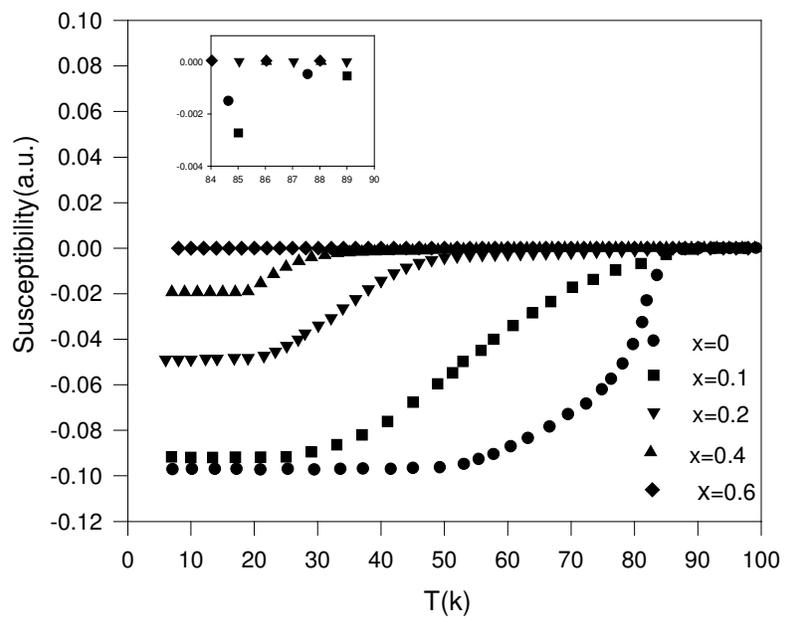

Fig.1

H.Salamati et al

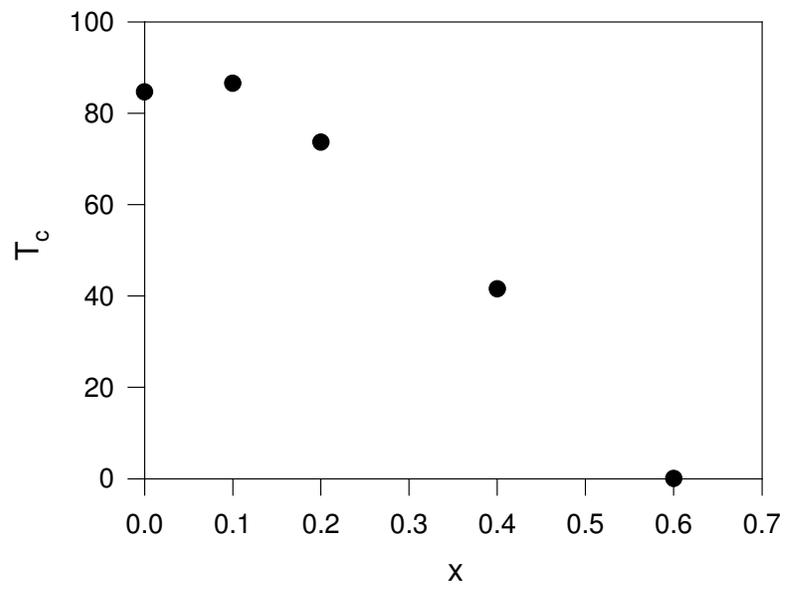

Fig.2

H.Salamati et al

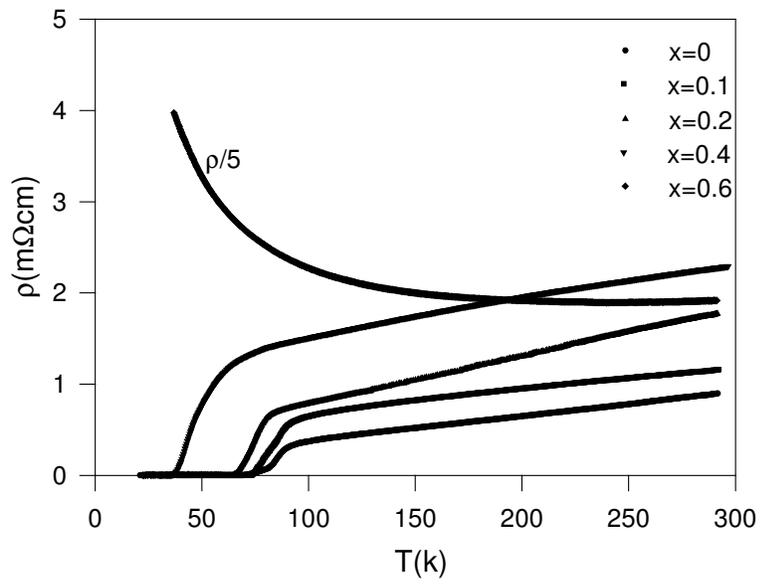

Fig.3

H.Salamati et al

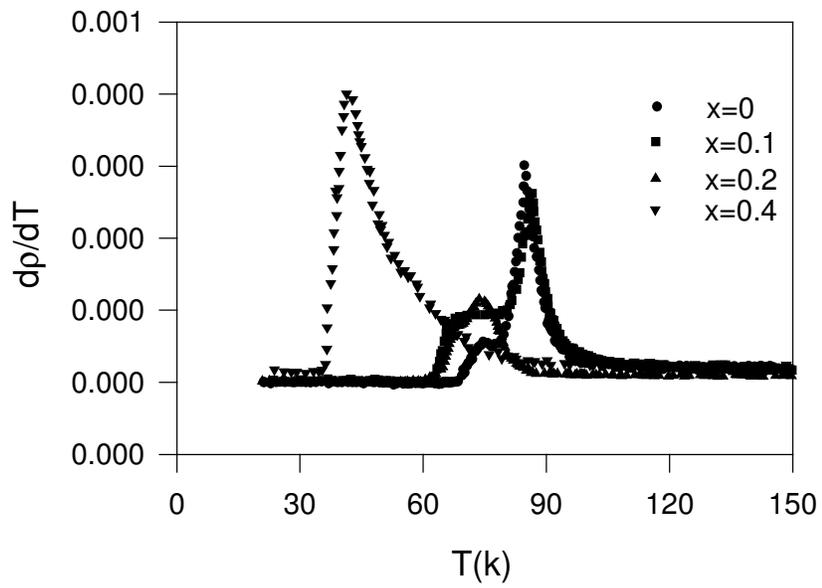

Fig.4

H.Salamati

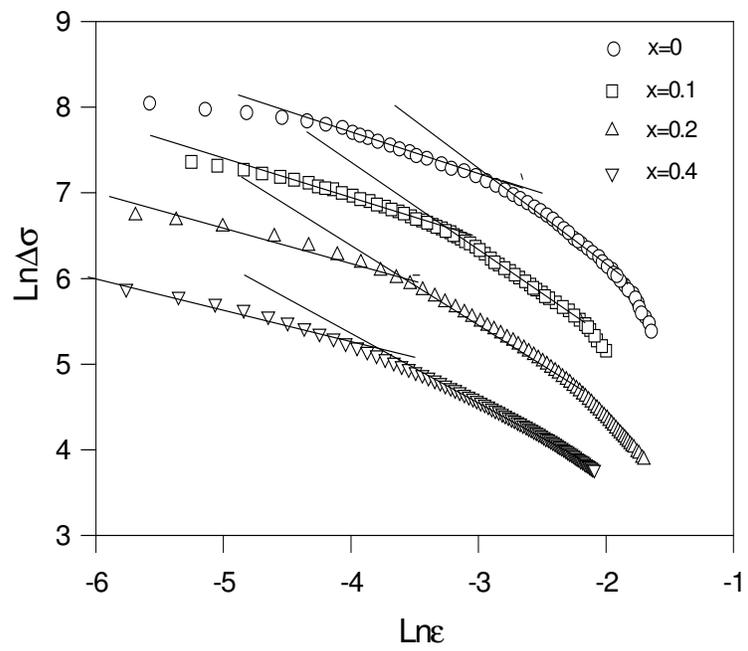

Fig.5

H.Salamati et al

| x | $T_c$ | $\lambda_1$ | $\lambda_2$ | $\varepsilon_0$ | J |
|---|---|---|---|---|---|
| 0 | 84.6 | -0.5 | -1 | 0.052 | 0.013 |
| 0.1 | 86.5 | -0.5 | -1 | 0.04 | 0.01 |
| 0.2 | 73.67 | -0.48 | -0.97 | 0.028 | 0.007 |
| 0.4 | 41.3 | -0.47 | -0.92 | 0.024 | 0.006 |
| 0.6 | 0 | | | | |

Table.1

H.Salamati et al